\documentstyle[prl,aps,epsf]{revtex}

\begin{document}
\draft

\twocolumn[\hsize\textwidth\columnwidth\hsize\csname @twocolumnfalse\endcsname

\title{Vortex Plastic Flow, $B(x,y,H(t))$, $M(H(t))$, $J_c(B(t))$,
Deep in the Bose Glass and Mott-Insulator Regimes}
\author{C.~Reichhardt, C.~J.~Olson, J.~Groth, Stuart Field, and Franco Nori}
\address{Department of Physics, The University of Michigan, Ann Arbor,
 Michigan 48109-1120}

\date{\today}
\maketitle
\begin{abstract}
We present simulations of flux-gradient-driven superconducting vortices
interacting with strong
columnar pinning defects as an external field $H(t)$
is quasi-statically swept from zero through a matching field $B_{\phi}$.
We analyze several measurable quantities, including the local flux
density $ B(x,y,H(t))$, magnetization $ M(H(t))$, critical current
$ J_{c}(B(t))$, and the  individual vortex flow paths.
We find a significant
change in the behavior of these quantities as the local
flux density crosses  $B_{\phi}$, and quantify it for many microscopic
pinning parameters.  Further, we
find that for a given pin density $J_c(B)$ can be enhanced
by maximizing the distance between the pins for $ B < B_{\phi} $.
\end{abstract}


\vskip2pc]
\narrowtext

{\em Introduction.---} Flux pinning by correlated disorder
in superconductors has recently been the subject of intense
investigations because of its ability to immobilize vortices,
reduce dissipation effects, and create high
critical currents \cite{reviews}.
In particular, the introduction of correlated disorder by heavy ion
irradiation can  create columnar defects in
which vortices are pinned along their entire length \cite{Civale}.
Such systems are predicted to form a Bose glass at low temperatures
when the vortices  localize at randomly-placed columnar pins
\cite{Nelson,Tauber}. The behavior of this state  depends on whether
the  magnetic flux density $B$ is below, equal to, or greater than
$B_{\phi}$, the matching field at which
the number of flux lines equals the number of columnar
 pins \cite{Shapiro,Leo,Rosenbaum}.
 In addition, an important role is played by the critical
state \cite{Bean,Kim}, in which a gradient in the flux profile arises as
individual vortices enter the sample and become pinned at defects.
This gradient determines $ J_{c}(B)$ \cite{reviews,Bean,Kim}.
Detailed information on the vortex dynamics in the critical state,
including {\it local\/} and averaged physical quantities,
is required to explain the
 effects induced by the addition of columnar defects. Furthermore,
understanding the effect of the spatial distribution of the pinning sites
on the Bose glass can facilitate
creating  samples with optimal pinning to enhance $J_c$,
which is of technological importance.
To investigate the  {\it flux-gradient-driven} dynamics of
vortices in superconductors with columnar defects, we perform simulations of
individual vortices {\it and\/} antivortices interacting with
strong columnar pinning sites, as an external field $H(t)$ is
swept through a complete loop.
We monitor distinct changes in relevant physical quantities
(e.g., $B$, $M$, $J_{c}$).

{\em Simulation.---} We simulate an infinite slab with a magnetic field
$ {\bf H} = H {\bf z} $  applied parallel to the surface so that there
are no demagnetization effects.
We consider rigid vortices and straight columnar pins of uniform
strength (all $\parallel {\bf z}$); thus we only
need to model a transverse 2D slice ($x$--$y$ plane) from the 3D slab.
Our results are for a  $36 \lambda \times 36 \lambda $ system
 with periodic boundary conditions,
 where $ \lambda $ is the penetration depth.
Our superconducting system has a pinned region from
$ x = 6\lambda $ to $ x = 30\lambda $; thus $1/3$ of the system
is unpinned and $2/3$ has randomly-placed non-overlapping parabolic traps
with radius $\xi_p = 0.15 \lambda$.
The flux lines evolve according to a $T=0$ molecular dynamics algorithm.
Thus, thermal effects are neglected and  we consider a situation deep in the
Bose glass regime, where the Mott insulator phase should be observable
\cite{Nelson,Tauber,Leo}.
Note that pinning due to columnar disorder is much less sensitive to thermal
effects than pinning due to point defects. For instance, for
$B < B_{\phi}$ T\"auber {\it et al.}\cite{Tauber}, working
with Bi$_{2}$Sr$_{2}$CaCu$_{2}$0$_8$, find that thermal effects on pinning
become relevant only for
$ T_1 \approx 0.9 T_c \approx 78 $ K $ \approx T_{irrev}$.
Therefore, in this case all temperatures $ T < T_1 $ may be considered low.

We simulate an increasing external field $H(t)$ as described in
\cite{Richardson,Reichhardt} where we quasi-statically add
flux lines to the unpinned region; these flux lines attain a uniform
density $n$, so that an external applied field $H$ may be defined as
$H=n\Phi_{0}$. As the external field increases, the flux
lines---by their own mutual repulsion---are forced into the pinned region
where their motion is impeded by the defects. We correctly model the
vortex-vortex force by a modified Bessel function $ K_{1}(r/\lambda) $,
which falls off exponentially so that a cutoff at $ 6\lambda $ can be
safely imposed. The overdamped equation of motion is:
${\bf f}_{i} = {\bf f}_{i}^{vv} + {\bf f}_{i}^{vp} = \eta{\bf v}_{i}$,
where the total force ${\bf f}_{i}$ on vortex $i$ (due to other vortices
$ {\bf f}_{i}^{vv} $, and  pinning sites  $ {\bf f}_{i}^{vp} $) is
$
 { \bf f}_{i}  =   \ \sum_{j=1}^{N_{v}}\, f_{0} \,
  K_{1}      (        |{\bf r}_{i} - {\bf r}_{j}|  / \lambda        )
  \, {\bf {\hat r}}_{ij}
   + \sum_{k=1}^{N_{p}}
 \frac{f_{p}}{\xi_{p}}
 \ |{\bf r}_{i} - {\bf r}_{k}^{(p)}|\ \
  \Theta ( ( \xi_{p} - |{\bf r}_{i} - {\bf r}_{k}^{(p)}| ) / \lambda )
  \ {\bf {\hat r}}_{ik} \
$.
Here $\Theta $ is the Heaviside step function,
$ {\bf r}_{i} $ ($ {\bf v}_{i}) $ is the location (velocity) of
the $i$th vortex,
%
%
%
$ {\bf r}_{k}^{(p)} $ is the location of the $k$th pinning site,
$ \xi_{p}$ is the radius of the pinning well, $ f_{p}$ is the maximum
pinning force, and we take $\eta=1$.
We measure lengths in units of the penetration depth $\lambda$,
forces in terms of $ f_{0} = \Phi_{0}^{2}/8\pi^{2}\lambda^{3}$
($=2\epsilon_0/\lambda$),
energies in units of $\epsilon_0=(\Phi_0/4\pi\lambda)^2$ ($=f_0\lambda/2$),
 and magnetic fields in units of $\Phi_0/\lambda^2$. The sign of the
interaction between vortices is determined by $ f_{v} $; we take
$ f_{v} = +f_{0} $ for repulsive vortex-vortex interactions and
$ f_{v} = -f_{0} $ for attractive vortex-antivortex interactions. A
vortex and antivortex annihilate and are removed from the system if they
come within $ 0.3\lambda $ of one another.
For  columnar pinning  the condensation energy is
 $ \epsilon_{0} = (\Phi_{0}/4\pi\lambda)^{2} $, which can be
related to  the maximum pinning force $f_p$  to give
 $ f_p/f_0 = \lambda/\xi_{p} = \lambda/0.15\lambda \approx 7 $ .
However the measurable quantities of the system do not change
for pinning forces  greater than $ 2.0f_{0}$.
We systematically vary one parameter such as the
density of pins, $ n_{p}$, pinning strength $f_{p}$, or the spatial
distribution of pinning, while keeping the other parameters fixed.
Previous simulations of flux-gradient-driven vortices
\cite{Richardson,Reichhardt} considered situations where $B < B_{\phi}$
and $f_{p}/f_{0} < 1$ and found a $J_{c}(H)$ dependence as $1/H$.
This corresponds to the experimentally common situation described by the
Kim model\cite{Kim}. However, when we consider very strong columnar pins
($f_p/f_0 > 2.0$) we find a very different behavior.

{\em Flux Profiles.---} From the location of the vortices,
we compute their local density
$B(x,y,H(t))$ and profile $B(x,H(t))$ inside the sample.
Figure 1 shows a typical example of $B(x)$ for a hysteresis loop.
When $ |B| < B_{\phi}$, the gradient of the flux profile
(i.e., the current $J_c$) is very large. Above $ B_{\phi} $
we see an abrupt change  in the flux profile to a much shallower slope.
As the external field is ramped still higher, the slope of $B$
decreases gradually.
These results are very similar to those observed experimentally
\cite{Rosenbaum,Majer} for samples with columnar pins as the local
field is increased beyond $B_{\phi}$
and quite different from the ones seen in other types of samples
(e.g., \cite{Richardson,Reichhardt}).
Such results have been interpreted in terms of a change in the pinning force:
when $B < B_{\phi}$ the vortices are pinned at individual defects, and
when $B > B_{\phi}$ many vortices are pinned at interstitial sites
due to the repulsion from the defect-pinned vortices.

To give a more microscopic idea of the behavior in this system,
in Fig.~2(a--c) we present snapshots of the positions of the
pins and the vortices in an  $ 8\lambda\times8\lambda $ region of the
sample (while the external field is gradually increased).
The vortices are entering from the left of the figures and thus
there is a flux gradient in the three frames.
In Fig.~2(a), most of the vortices are pinned at defects;
only two (located near other defect-pinned flux lines) are at
interstitial sites.
Since most of the vortices at these low fields are trapped by defects,
the effective pinning force is very high and hence, as seen in Fig.~1,
the flux gradient for low fields is quite steep.
In Fig.~2(b), near $ B_{\phi} $, a greater fraction of interstitially
pinned vortices is present, producing a much lower flux gradient.
The fact that unoccupied pins are typically located very close to occupied
pins indicates that many unoccupied pins are being ``screened" by the vortex
repulsion from the occupied pins.
Also, because of the flux gradient, there are more interstitially pinned
vortices near the left of the figure than the right.
Thus, the presence of a flux gradient limits the number of
accessible pinning sites, when these sites are randomly distributed.
Finally in Fig.~2(c), above  $ B_{\phi} $,  the flux gradient is small
and hence the effective pinning force is low.
Here a large number of vortices are weakly trapped at interstitial sites and
the majority of the defects are now occupied by vortices.
In the lower left corner a small domain of triangularly-arranged vortices
can be seen---indicating that the vortex-vortex repulsion is starting to
dominate over the vortex-pin attraction.

{\em Magnetization loops and Critical Current.---} From the microscopic
vortex dynamics of individual vortices described above,
we can compute macroscopic measurable quantities.
In particular, we quantify the vortex behavior described above,
as the field is brought through $ B_{\phi} $,
by measuring $ M(H) $ and  $ J_{c}(B) $
for samples with different pinning parameters (see Fig.~(3)).
For instance, the effect of changing the matching field $ B_{\phi} $
is examined in Fig.~3(a,b) by fixing $ f_{p} $ at $ 2.5f_{0}$
and changing the pin density $n_p$.
For all the cases we considered,
the width of the magnetization curve is much broader when
$ | B | < B_{\phi} $, falling off very rapidly when $ B  >  B_{\phi} $.
Fig.~3 shows the enhancement of $J_{c}(B)$, for $B < B_{\phi}$,
as the field is swept.
We  obtain $ J_{c}(B) $  directly from the flux density profiles
using Maxwell's equation $ dB/dx = \mu_{0}J $.
For a fixed field $H$, we average the slope of $B$ over the entire sample.
In all cases the enhancement of $ J_{c}(B) $ is restricted to
$ | B | < B_{\phi} $.
This  is very similar to the results seen in \cite{Majer} at low $ T $,
where  $ J_{c}(B) $ was also obtained directly from the flux gradient.
It can also be seen from Fig.~3 that $J_{c}$ has a definite
dependence on $B$, which is different from the step-like two-current model
suggested in \cite{Rosenbaum} (indicated by a dashed line in Fig.~3(b)).

We next examine the effect of varying the pinning force from
$ f_{p} = 0.3f_{0} $ to $f_{p} = 5.0f_{0}$ (see Fig.~3(c,d)).
There is little change in $M(H)$ and $J_c(B)$ as the pinning force
is increased from $ f_{p} = 2.5f_{0}$ to $ f_{p} = 5.0f_{0}$.
This reflects the fact that once vortices are trapped strongly enough,
so that they cannot be unpinned, any further increase in the pinning
strength does not significantly affect $M(H)$ and $J_c(B)$.
As the pinning force is decreased in Fig.~3(c), the width of the
magnetization loop narrows. A crossover at $B_{\phi}$ is no longer
observed when the pinning force from the individual defects becomes on the
order of the interstitial pinning force (when $f_p \sim f_0/2$).
This behavior is emphasized in  Fig.~3(d) for $ J_{c}(B) $.
At the lowest pinning strengths, only a small increase in $ J_{c}(B) $
can be seen when $|B|<B_{\phi}/3$, and not when $|B|<B_{\phi}$ as in the
strongly pinned ($f_p \gtrsim 2.0 f_0$) cases.
These results are very similar to those  seen in \cite{Majer}
when the temperature was raised,
which reduces the effective pinning strength.

The theory of a Mott insulator phase in the Bose glass \cite{Nelson}
leads one to expect an enhancement in $ J_{c}(B) $  at $ B_{\phi} $.
At this field the vortices become strongly localized at
the pinning sites, due to the repulsion of the other vortices,
and produce a Meissner-like phase which prevents
increased flux penetration over a range of field.
However, we  do {\it not\/} observe any particular enhancement of
$ J_{c} $ at $ B = B_{\phi} $.
This is due to the combined effects of {\it both\/}
the critical state, which prevents a uniform field in the sample,
and the randomness in the spatial distribution of defects,
which prevents all the pins from being occupied due to ``screening"
effects.  We point out that near $B_{\phi}$ there can still be a
decrease in the  magnetization relaxation rate
since it depends on thermal activation and tunneling.

We can significantly reduce screening effects (due to closely-spaced defects)
by changing the spatial arrangement of the pinning sites, thus
enhancing $J_{c}$.
Figure 3(e,f) presents $M(H)$ and $J_c(B)$ for three systems with
pinning sites which are
(1) randomly located,
(2) placed in a triangular lattice (see, e.g., \cite{Baert}), and
(3) randomly displaced, up to $0.25 \lambda$ from a triangular lattice.
All three systems have
$ B_{\phi} = 1/\lambda^{2}$ and $ f_{p} = 2.5f_{0} $.
For fields less then $ B_{\phi} $, the width of $M(H)$ and $J_{c}(B)$
for the triangular pinning is nearly twice that for the
random case.  The asymmetry in  $ M(H) $ and $ J_{c}(B) $ is much more
pronounced when pins are placed in a triangular lattice.
For this case, there is also a very sharp drop in $ M(H) $
just past $ B_{\phi} $, very similar to that observed in \cite{Baert},
since here the distinction between vortex pinning at defects and
interstitial pinning is much better defined.
This result is very important since it indicates that substantial
increases in  the critical current for fields less then
$ B_{\phi} $ can be achieved by evenly spacing the pins.  In this case,
$M(H)$ and $ J_{c}(B) $ are enhanced because all the pins are
accessible to the vortices, as opposed to the random case where
it is difficult for vortices to reach closely-spaced defects.
Moreover, in the  system with triangular pinning, trapped vortices
are better localized at the defects due to the interactions with
other pinned vortices (since an additional energy minima
is present when the vortices are in a favorable lattice configuration).
This indicates that the Mott insulator phase is more accessible in a
triangular array of pinning sites than in a Bose glass with
randommly-located defects.
In the case where the pinning sites are randomly displaced
(up to $0.25\lambda$) from the triangular sites,
there is still a clear enhancement of $ J_{c}(B) $; however, it is much
smaller than in the ordered triangular pinning array.

{\em Vortex Plastic Flow.---}
To investigate the dynamics of the flux-gradient-driven vortices
near $B_{\phi}$ and higher fields, we present in Fig.~4(a,b)
the trajectories of the vortices for a system with (a) strong pinning,
$ f_{p} = 2.5f_{0} $, and with (b) weaker pinning,
$ f_{p} = 0.3f_{0} $, both with $B_{\phi} = 1.0\Phi_{0}/\lambda^{2}$.
For both panels the external field $H$ is increased from
$ 0.9\Phi_{0}/\lambda^{2} $ to $ 1.4\Phi_{0}/\lambda^{2} $.
In (a) the moving vortices follow specific winding channels
indicative of { \it plastic flow\/} of interstitial vortices
around domains of flux lines which are strongly pinned at defects.
These ``vortex rivers" often develop well defined tight {\it bottlenecks\/}
as well as much broader ``vortex streets" with several lanes.
At the top left side of Fig.~4(a), two bottlenecks can be seen.
Each of these ``interstitial vortex paths" is located between
flux lines that are strongly pinned at defects,
indicating that the  interstitial vortices are flowing through
the energy minima created by the strongly pinned flux lines.
These  interstitial-vortex-channels form for both the entrance and exit
of vortices and are a general feature of all simulations with randomly
placed strong pinning sites.
It is interesting to note that although a large number of vortices
are clearly immobile at the defects,  $J_{c}$ is quite
low for this region of $B > B_{\phi}$ indicating that the channels are
acting as {\it weak links}.
In Fig.~3(b) with  $ f_{p} = 0.3f_{0} $ we see a much different behavior
with only a few channels. Further, it can be observed from the lines
that cross through pinning sites that vortices can
become unpinned since the pinning force is on the order of the interstitial
pinning. Thus vortex motion consists of both interstitial trajectories
and also pin-to-pin motion.

{\it Conclusions.---}
Our simulations of flux-gradient-driven rigid vortices interacting
with  columnar pinning sites are consistent with recent experiments
\cite{Rosenbaum}
which indicate that a sharp change in the magnetic flux gradient
occurs above $B_{\phi}$.
Our results quantitatively predict how $ J_{c} $ varies with
field and pinning parameters (e.g., strength and location).
Moreover, we can monitor the spatio-temporal dynamics of vortices
as the pinning mechanism evolves from strong columnar pinning
(for $B<B_{\phi}$) to weaker interstitial pinning (for $B>B_{\phi}$).
In particular, we quantify the enhancement of $J_{c}$, when $B < B_{\phi}$,
by varying the spatial arrangements of the pinning sites.
When the pinning strength of the columnar defects becomes on the
order of the interstitial pinning strength,
this sharp transition in $J_{c}$ is greatly reduced.
We compute the spatio-temporal dynamics of vortices and show that
for strongly pinned samples the vortex transport is dominated by
the plastic flow of interstitially-pinned vortices around
regions of flux lines strongly pinned at defects.  In the weaker pinning
samples, vortices can jump from defect to defect site.
Finally, $ M $ and $ J_{c} $ can  be considerably enhanced by placing
the defects on a regular triangular lattice, for a given pin density,
so that all pins are equally accessible, thus preventing the screening
of closely-spaced defects occurring in a random distribution.

This work was supported in part by the NSF under grant No.~DMR-92-22541.
CR and CJO acknowledge support from Rackham graduate fellowships.
We also acknowledge the UM Center for Parallel Computing for computer time,
and J.~Siegel for a critical reading of this manuscript.


\vspace{-0.25in}

\begin{figure}
\centerline{
\epsfxsize=3.4in
\epsfbox{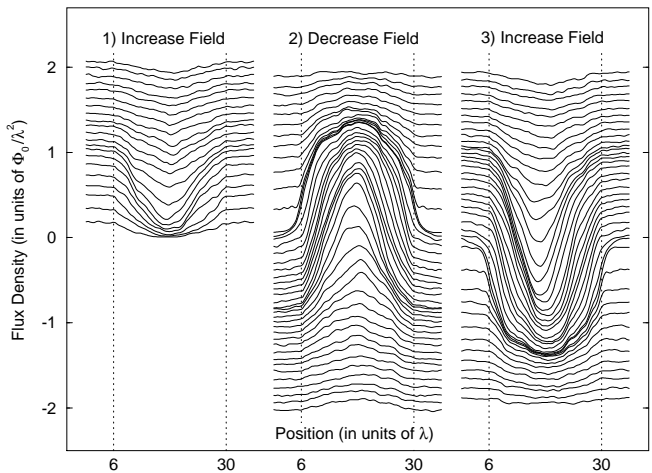}}
\vspace{0.05in}
\protect\caption{
\sloppy {
\mbox{Magnetic flux density profiles,
$B(x,H(t))$ $=$}
$(36\lambda)^{-1} $$\int_{0}^{36\lambda} $$ d\!y B(x,y,H(t))$,
for a $ 36 \lambda \times 36 \lambda $ system with a
$24 \lambda \times 36 \lambda $ region (the actual sample) containing
864 pinning sites with $f_{p}$$ =$$ 2.5f_{0}$, $\xi_p $$=$$ 0.15\lambda$.
Here
$B_{\phi} $$=$$ 1.0 \, {\Phi}_0/{\lambda}^2 $$=$$
864 \Phi_{0} / (36 \lambda \times 24\lambda)$.
For the initial ramp-up phase in (1) a total of $2600$ vortices are added,
so the maximum external field is
$2.0 \, {\Phi}_0/{\lambda}^2 $$\approx $$ 2600\Phi_{0}/(36\lambda
\times 36\lambda)$.
In (2) the field is ramped down and then reversed to a final value of
$-2.0 \, {\Phi}_0/{\lambda}^2$. Finally in (3), the
field is brought up to $2.0 \, {\Phi}_0/{\lambda}^2$.
A large gradient in $B$ can be seen for $|B| < B_{\phi}$.
}
}
\label{fig1}\end{figure}

\vspace{-0.25in}

\begin{figure}
\centerline{
\epsfxsize=3.4in
\epsfbox{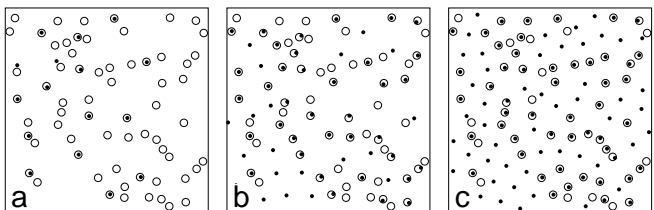}}
\vspace{0.05in}
\protect\caption{
An  $8\lambda \times 8\lambda$ region  of the
$24\lambda \times 36\lambda$ sample studied in Fig.~1.
The vortices enter from the left edge of each frame
(which corresponds to the sample edge).
Panels (a), (b), and (c) correspond
to $ B/B_{\phi} \approx 0.3$, $1.0$, and $ 1.6 $,
respectively, of the ramp up phase in Fig.~1.
Pinning sites are indicated by open circles, while vortices are
shown as filled dots.
}
\label{fig2}\end{figure}

\vspace{-0.25in}

\begin{figure}
\centerline{
\epsfxsize=3.4in
\epsfbox{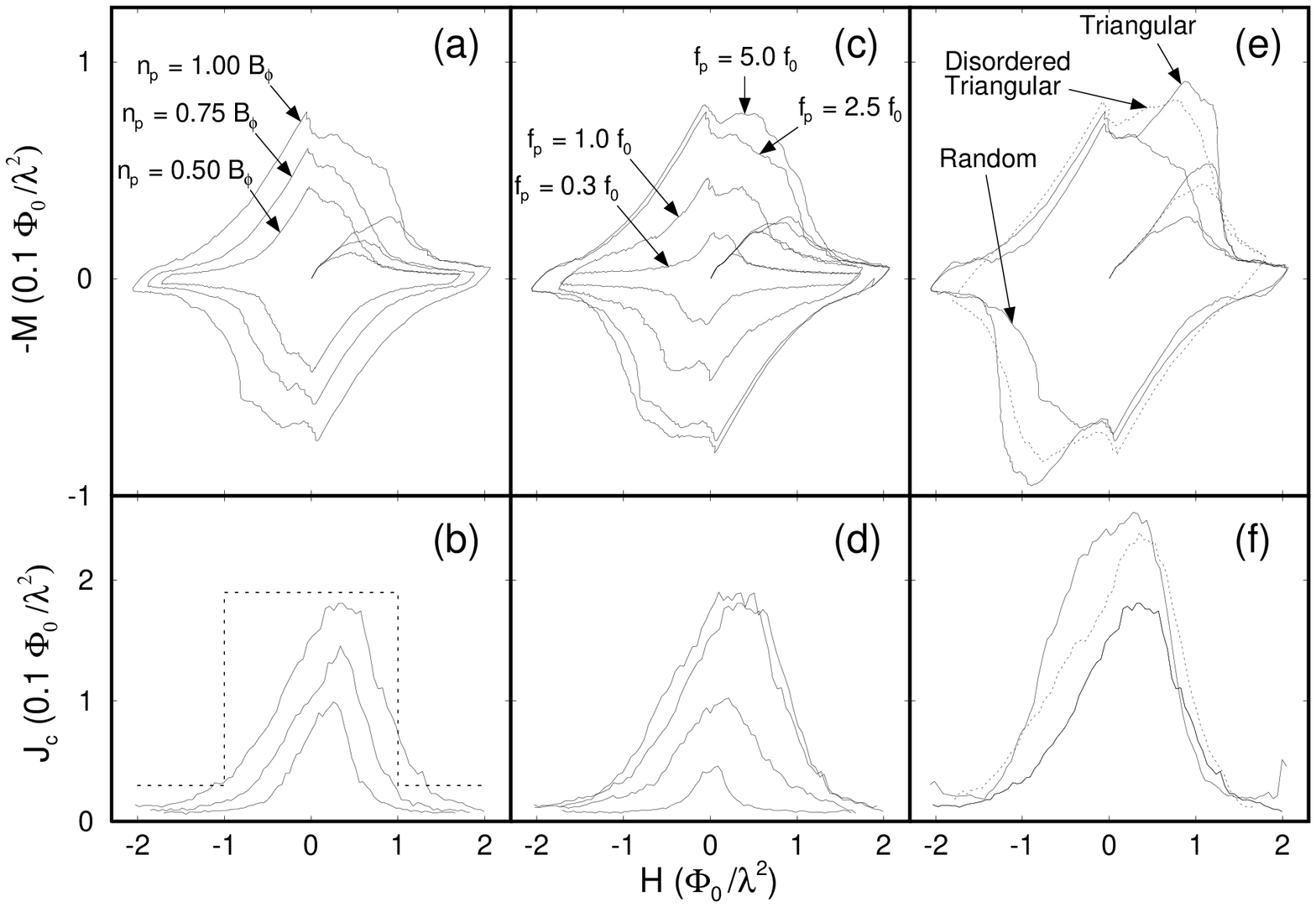}}
\vspace{0.05in}
\protect\caption{
Magnetization loops (top panels) and the corresponding
critical currents (bottom panels) for several samples.
The $J_c(B)$s are taken directly from the $B(x)$ during ``ramp-down"
(e.g., stage (2) in Fig.~1).
%
%
%
In (a,b) $f_p$ is held fixed at $2.5f_0$ and
the density of pinning sites $n_p$ is varied:
$B_{\phi}/\Phi_0 = 0.50/{\lambda}^2$, $0.75/{\lambda}^2$, $1.0/{\lambda}^2$.
In (c,d) $n_p$ remains fixed ($B_{\phi} = 1.0 \Phi_0 /{\lambda}^2$)
and the pinning strength $f_p$ is changed.
In (e,f) $B_{\phi} = 1.0 \Phi_0 /{\lambda}^2$, $f_p = 2.5f_0$, and the
location of the pinning sites is varied.
(f) shows the significant {\it enhancement} of $J_c(B)$ that results
from defects placed in a regular triangular array, as opposed to
random placement.
These results show that even a distorted triangular array of pinning
sites significantly enhances $J_c(B)$ over the case with a random
location of pinning sites.
}
\label{fig3}\end{figure}

\vskip -0.25in

\begin{figure}
\centerline{
\epsfxsize=3.4in
\epsfbox{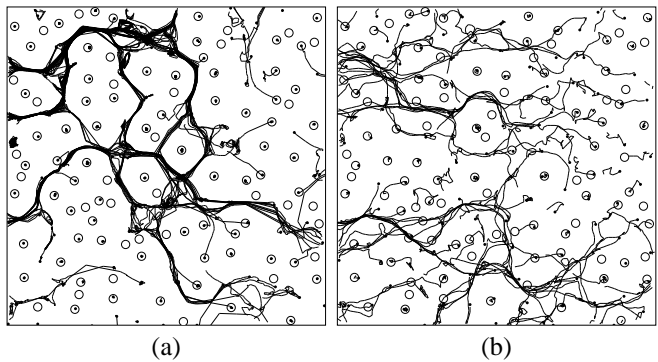}}
\vspace{0.05in}
\protect\caption{
(a,b) show the trajectories of vortices while the external
field is raised from
 $ 0.95 \, \Phi_{0}/\lambda^{2}$ to $1.4 \, \Phi_{0}/\lambda^{2}$.
(a) A $10\lambda\times10\lambda$ region of the sample used in Fig.~1.
The strength of the pinning force, $f_{p}$, is $2.5f_{0}$,
for the strong-pinning case (a), and $0.3f_{0}$,
for the weak-pinning case (b).
In (a) the vortex transport is characterized by vortex trails of
{\it interstitial\/} vortices which move around regions with
flux lines that are strongly-pinned at defects.
In (b) vortex transport proceeds in a different manner: pin-to-pin
vortex motion, as well as interstitial, is possible and the
previously-narrow vortex trails become considerably broader.
%
%
}
\label{fig4}\end{figure}

\end{document}